\documentclass[aip,jap,amsmath,amssymb,amstext,citeautoscript,punctuation,reprint]{revtex4-1}

\usepackage{graphicx,xcolor,charter}

\begin{document}

\sloppy


\title{Influencing the Martensitic Phase Transformation in NiTi through Point Defects}

\author{A. Mansouri Tehrani}
\affiliation{Department of Chemistry, University of Houston, Houston, TX 77204 USA}
\author{H. Shahrokhshahi}
\affiliation{Department of Mining and Metallurgical Engineering, Amirkabir University of Technology, Tehran, Iran}
\author{N. Parvin}
\affiliation{Department of Mining and Metallurgical Engineering, Amirkabir University of Technology, Tehran, Iran}
\author{J. Brgoch*}
\affiliation{Department of Chemistry, University of Houston, Houston, TX 77204 USA}
\email{jbrgoch@uh.edu}

\date{\today}

\begin{abstract}

Equiatomic nickle-titanium (NiTi) is investigated to determine the consequences of point defects on the Martensitic phase transformation.  Using molecular dynamics (MD) simulations, NiTi with 0.1\%, 0.5\%, 1\%, 2\%, and 3\% of Schottky-type defects (vacancies) have been modeled with the temperature of structural transformation elucidated. Increasing the concentration of point defects leads to this transformation occurring at lower temperatures than the perfect structure while the final monoclinic unit cell angle ($\gamma$) substantially decreases.  Modeling anti-site defects at the 0.1\%, 0.5\%, and 1\% concentration level indicates the cubic to monoclinic structural transformation temperature decreases even faster with a more dramatic change in $\gamma$ compared to the vacancy structure. The change in this Martensitic transformation stems from pinning due to the structural defects.
\end{abstract}


\maketitle 

\section{Introduction}

\begin{figure} 
\includegraphics[width=3in]{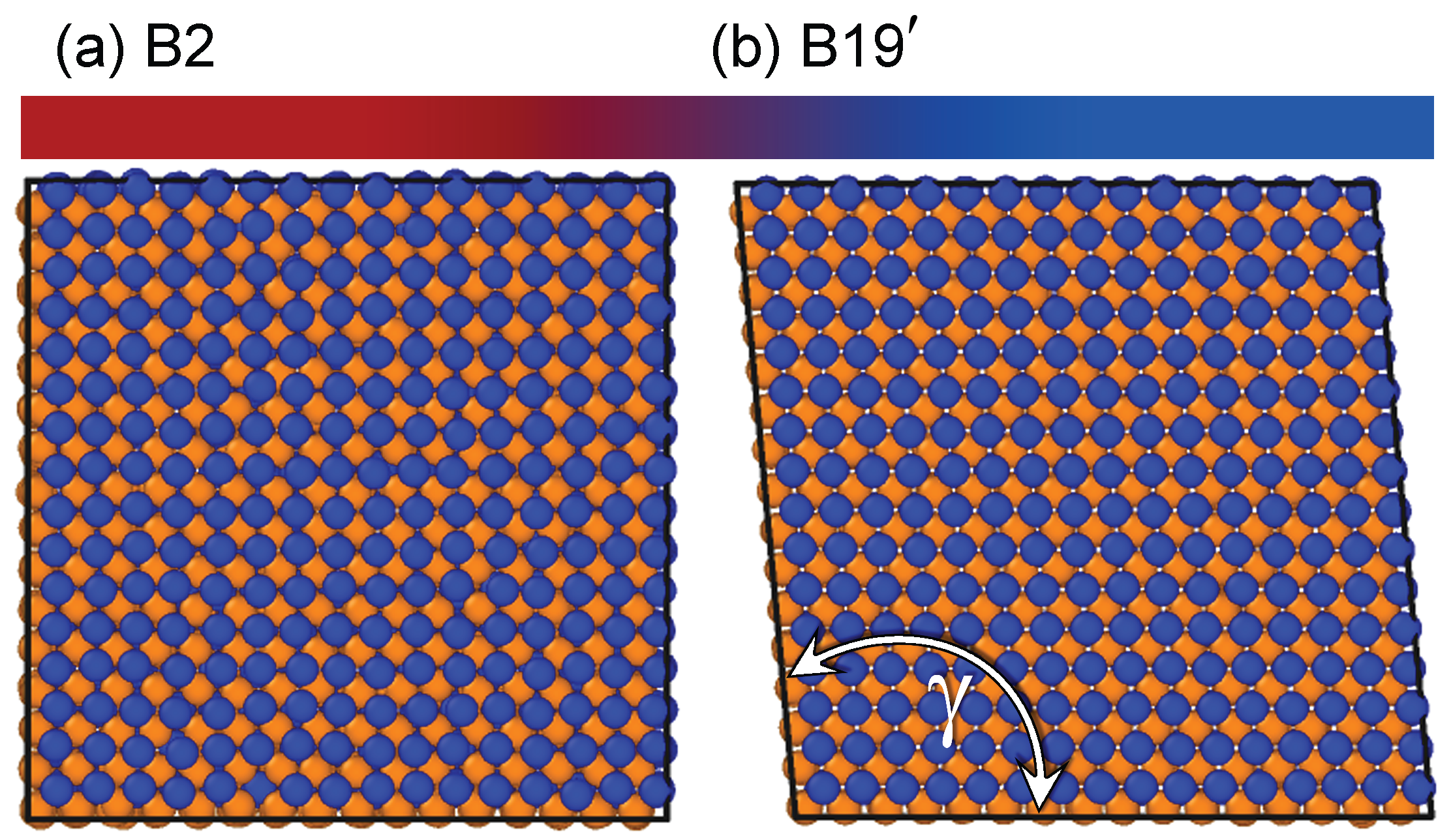} 
\caption{The structural transformation of 
NiTi from (a) cubic B2 structure to (b) monoclinic B19$^\prime$ structure with the angle, $\gamma$, highlighted} 
\label{fig:NiTi-structure}
\end{figure}

NiTi is an ideal material for multifarious practical application due to its reversible, diffusionless transformation between the austenite and Martensite phases.\cite{Otsuka1971} At higher temperature, NiTi forms the cubic B2 (CsCl-type) crystal structure while at lower temperature the compound forms the monoclinic B19$^\prime$ crystal structure, as shown in Fig.\,\ref{fig:NiTi-structure}.  This transformation is induced through external strain or changing temperature and has been extensively investigated through computation and experiment. 
For example, density functional (DFT) calculations \cite{Wagner2008, Hatcher2009, Holec2011}  
and molecular dynamics (MD) simulations\cite{Saitoh2006, Uehara2009} both accurately reproduce the stress-induced Martensitic transformation in NiTi.  The temperature-induced phase transformation has also been analyzed by tight-binding calculations and MD simulation.\cite{Mutter2010} 
On the experimental side, the temperature-induced transformations are often monitored \textit{via} differential scanning calorimetry (DSC). The shape memory and pseudelasticity properties of nearly equiatomic NiTi have been studied across a range of temperatures from $-$20$^\circ$C to 100$^\circ$C. It is shown that the loading rate as well as temperature could both have great effect on the shape memory behavior of NiTi. \cite{Shaw1995} 

Beyond developing a fundamental understanding of this structural transformation, research has also focused on the ability to control the transformation by varying the composition. Modifying the concentration of Ni and Ti substantially changes the transformation temperature. Experiments show that increasing Ni concentration leads to a decrease in Martensitic transformation temperature. \cite{Wang1965,Frenzel2010,Khalil-Allafi2002} This variation in transformation temperature has qualitatively been confirmed using MD simulations between 47\% to 53\% of Ni content.\cite{Mutter2010}
 
Nearly every previous report focuses specifically on the perfect, defect-free NiTi crystal structure.\cite{Huang2003,Sanati1998}  Yet, studies have shown Schottky-type defects (vacancies) as well as anti-site defects are both energetically favorable in NiTi and thus should be expected in the products.\cite{Lutton1990, Weber2014, Lu2007}  Anti-site defects have been previously modeled through MD simulation to understand the crystalline to amorphous transition of NiTi.\cite{Lai2000}  Through the random exchange of Ni atoms and Ti atoms in the structure, the crystalline to amorphous transition was reported to occur for a long-range order parameter (LRO) less than 0.4. This work clearly indicates the important relationship between chemical disorder and phase transition.  More recently, using a two-dimensional Lennard-Jones potential, the phase transformation of NiTi indicated the preferential formation of vacancies upon cycling.  These can act as nucleation sites for subsequent Martensitic transformations.\cite{Kastner2011} 

The work presented herein focuses on employing MD simulations to investigate the effects of point defect concentration on the temperature induced phase transformation in NiTi. Given the computational cost associated with DFT and the need for large supercells to capture the dilute concentration of defects ($<$3\%), MD simulations are the most viable computational method available.\cite{Wu2013}  Our work shows that increasing the concentration of point defects dramatically changes the temperature of the Martensitic phase transformation.  These calculations highlight the importance of defect concentration on the defect-property relationship in NiTi. Moreover, they provide a fundamental understanding of how point defects influence structural transformations.

\section{Model and Simulation Details}

Calculations were carried out using molecular dynamics method as implemented in the LAMMPS software package.\cite{Plimpton1995} The interatomic interactions were captured through the Finnis Sinclair type EAM (embedded atom method)\cite{Finnis1984} potential developed by Lai and Liu.\cite{Lai2000} Parameters for this potential have been fit from first principle calculations of the B2 phase at 0\,K.  Although simpler interatomic potentials, \textit{e.g.}, Lennard-Jones, have also been successfully used to reproduce the phase transformation of NiTi,\cite{Kastner2003} EAM type potentials possess a more accurate description of interatomic atomic interactions.\cite{Saitoh2006} Moreover, EAM potentials account for the many body effect and distinguish between different coordination environments, which is essential when vacancies are introduced into the system. For all calculations, periodic boundary conditions were implemented to reduce the finite size errors associated with the small simulation system. 

\begin{figure} 
\includegraphics[width=3in]{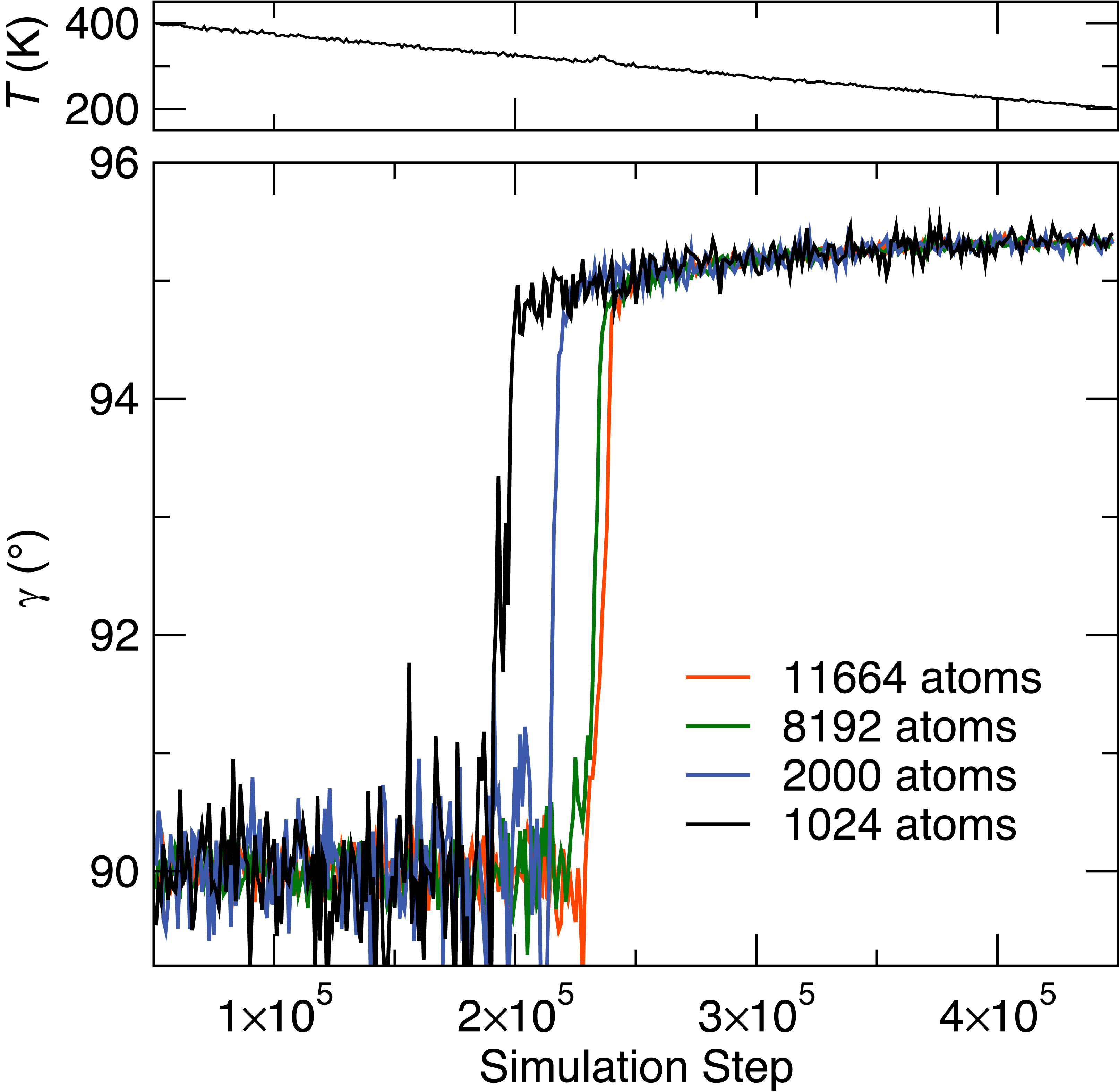} 
\caption{ The number of atoms modeled using MD for the cooling cycle of the NiTi Martensitic phase transformation  show finite size effects.  The bulk properties of NiTi are not reliable for systems smaller than 8192 atoms.} 
\label{fig:systemsize}
\end{figure}

To ensure the system's finite size does not affect the structural transformation, supercell calculations were repeated by examining various system sizes including of 1024 atoms (24\,\AA\,$\times$\,24\AA\,$\times$24\,\AA), 2000 atoms (30\,\AA\ $\times$ 30\AA\ $\times$ 30\,\AA),  8192 atoms (48\,\AA\,$\times$\,48\,\AA\,$\times$48\,\AA),  11664 atoms (54\,\AA\,$\times$\,54\AA\,$\times$54\,\AA),  and 16000 atoms  (60\,\AA\,$\times$\,60\AA\,$\times$60\,\AA). The transformation was determined by monitoring the unit cell angle ($\gamma$) to indicate the formation of the B19$^\prime$ structure. As illustrated in Fig.\,\ref{fig:systemsize}, the transformation temperature was highest for the 1024 atom system with a significant decrease in transformation temperature observed for the 2000 atom system.  Continuing to increase the system size to 8192 atoms further decreases the transformation temperature, which is then nearly constant with 11664 atoms.  Interestingly, modeling 16000 atoms shows this supercell is large enough to accommodate the formation of nano-twinning domains in accordance with the experimental identification.\cite{Xie2004}  The transformation temperature of this system is still nearly the same as the 11664 atom model.  As a result, the system with 8192 atoms was selected based on the compromise between accuracy and computational cost.

The initial configurations were relaxed at 400\,K and zero external pressure for 10\,ps, using a NPT ensemble. Nose-Hoover thermostat algorithm \cite{Hoover1985} and Parrinello-Rahman methods were used to control the temperature and pressure of the system respectively.  The time step was set at 0.5\,fs. The temperature dependent response of the system was examined by cooling following a ramp-rate of 1\,K per 2000 simulation steps until a system temperature of 200\,K was reached.  A heating cycle was then followed using the same ramp-rate to 400\,K. To investigate the effect of point defects, NiTi in the B2 structure was modeled by incorporating 0.1\%, 0.5\%, 1\%, 2\%, and 3\% percent of random Schottky-type defects ensuring the 1:1 stoichiometry was maintained.  The anti-site defects were also modeled as a random distribution of 0.1\%, 0.5\%, and 1\% disorder again making sure the elemental ratio was constant.  Visualization of the results was done by Ovito (Open Visualization Tool) software \cite{Stukowski2010}.

\section{Result and Discussion}

\subsection{Structural Transformation in Perfect NiTi}
The structural relaxation of the initial, perfect NiTi in the austenite crystal structure (B2) yields unit cell parameters at 400\,K, shown in Table 1, that are in excellent agreement with previous reported values.\cite{SLNNT2003} The relaxation of B19$^\prime$ at 200\,K also produces lattice parameters that differ by less than 3\% compared to literature. \cite{Huang2003,Lai2000} The greatest differences in the optimization of B19$^\prime$ arises from variation in $\gamma$, which is likely related to the interatomic potential used and the boundary conditions.  Nevertheless, the EAM used in this study is valid to describe this interatomic interactions\cite{Lai2000}.

\begin{table}
 \caption{The calculated NiTi lattice parameters for the B2 structure at 400\,K and B19$^\prime$ structure at 200\,K.  These values are compared against the previously reported experimental\cite{SLNNT2003} and calculated\cite{Lai2000,Huang2003} values.}
\begin{center}
 \begin{tabular} {l  l l l l} 

\hline \hline
~ & $a$ (\AA) & $b$ (\AA) & $c$ (\AA) & $\gamma$ ($^\circ$) \\ [0.5ex] 
 \hline
 B2 & 3.011 & 4.258 & 4.258 & 90.00 \\ 
 
  Literature (B2) \cite{SLNNT2003} & 3.013 & 4.261 & 4.261 & 90.00 \\
 
 B19$^\prime$ & 3.032 & 4.616 & 4.224 & 95.30 \\
 
  Literature (B19$^\prime$) \cite{Lai2000} & 2.956 & 4.455 & 4.189 & 93.26 \\
 
 Literature (B19$^\prime$) \cite{Huang2003} & 2.929 & 4.686 & 4.4048 & 97.80 \\
 \hline\hline
\end{tabular}
\end{center}
\end{table}

The structural transformation temperature is a critical component of shape memory alloys.  The temperature of the phase change is often described by the starting (onset) temperature as well as the temperature where the transformation is complete, \textit{i.e.}, the final temperature.  Using this description, the transformation of perfect NiTi is calculated to occur at $Ms$ = 310\,K, $Mf$ = 305\,K, $As$ = 339\,K, and $Af$ =346\,K, where $M$ is Martensite, $A$ is austenite, $s$ is the start (onset) temperature, and $f$ final temperature.  These calculated values agree well with the  reported phase transformation temperature.\cite{Saitoh2006, Pelton2000, Zhong2012, Ishida2007} This transformation also leads to the technologically important hysteretic behavior, which can be derived from cooling-heating cycle.  Here, the hysteresis is calculated by taking the difference between the maximum values of the derivatives for the transformation (Fig.\,\ref{fig:systemsize}).  In perfect NiTi, the hysteresis is 35\,K for the 8192 atom system.  This is in agreement with the previously calculated values, which range between 20\,K and 30\,K.\cite{Khalil-Allafi2009}

\subsection{The Influence of Point Defects}

\begin{figure}[b!] 
\includegraphics[width=3in]{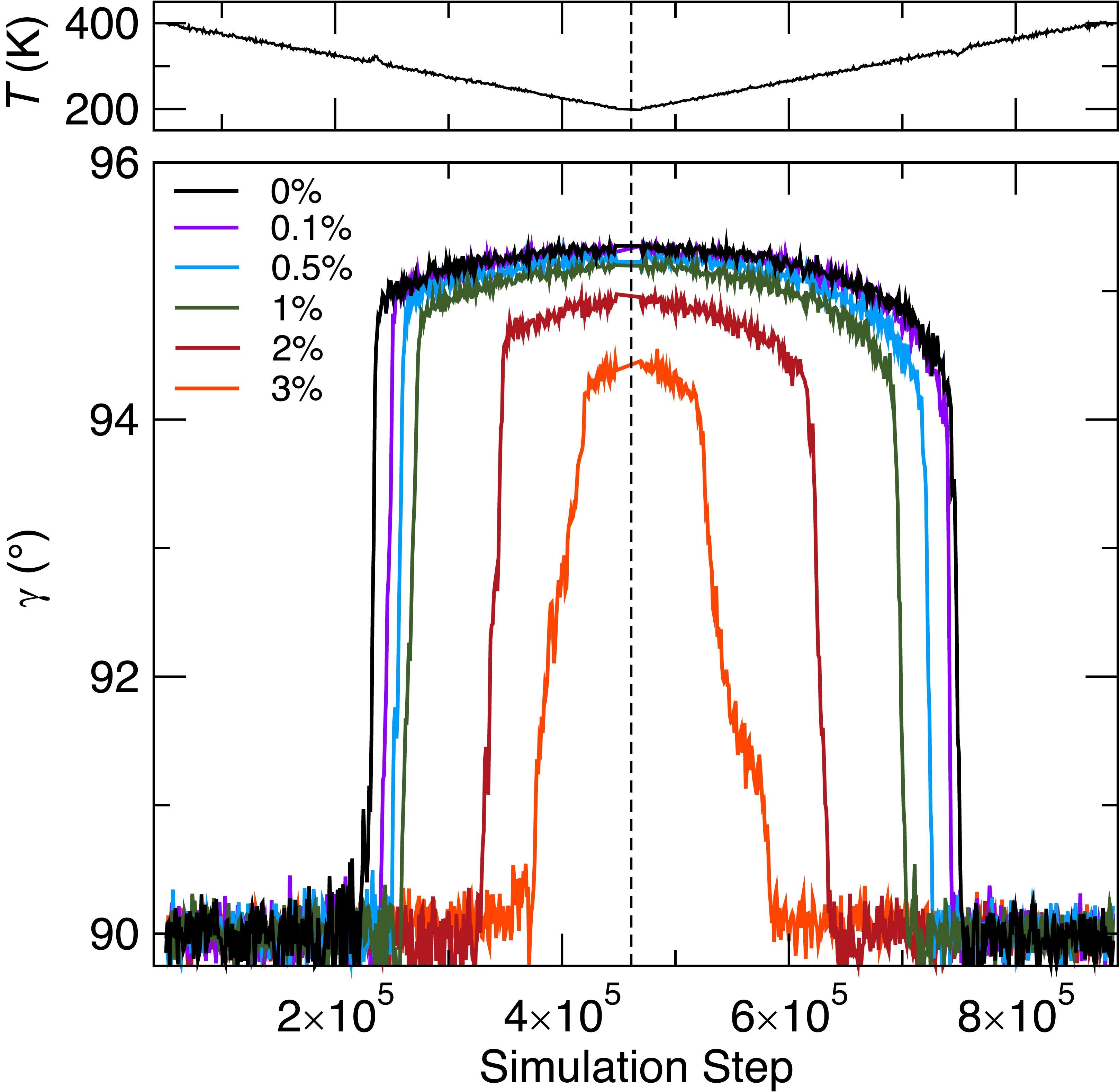} 
\caption{The change of unit cell angle ($\gamma$) with simulation step and temperature (shown above) with 0\%, 0.1\%, 0.5\%, 1\%, 2\%, and 3\% of Schottky-type defects (vacancies) indicates the structural transformation. The abrupt change in $\gamma$ correspond to the occurrence of phase change.} 
\label{fig:NiTi-degree-v}
\end{figure}

To understand the effects of chemical disorder on Martensitic transformation, the concentration of point defects was varied.  Specifically, 0.1\%, 0.5\%, 1\%, 2\%, and 3\% of Schottky-type vacancies were modeled as well as 0.1\%, 0.5\%, and 1\% of anti-site defects.  Although these concentrations are higher than the reported thermal equilibrium values,\cite{Lutton1990} choosing higher concentrations enables the effect of defects to be accentuated. It should be noted, however, that the synthesis procedure, especially when conducted at elevated temperatures, could yield these higher defect concentrations. Fig.\,\ref{fig:NiTi-degree-v} demonstrates the changes in $\gamma$ with respect to simulation step during cooling and heating cycles with different concentrations of vacancies. The abrupt change of $\gamma$ is attributed to the occurrence of the phase transformation.

A few notable features arise from the inclusion of random vacancies.  Increasing their concentration leads to a significant change in the temperature of structural transformation.  In all models, the increasing concentration of vacancies causes the transformation to occur at lower temperatures during the cooling cycle and the heating cycle.  Moreover, a suppression of $\gamma$ is present at higher vacancy concentrations. This transformation is also calculated to occur over a wider temperature range indicated by the slope of the transformation. These changes are most evident in NiTi with 3\% vacancies. 

The Martensitic transformations can be defined based on shear and dilatational displacements, the former being parallel and the latter normal to the habit plane.\cite{Patel1953}  Hence, this decrease in the temperature range and the supression of $\gamma$ is due to the vacancies inhibiting the transformation by reducing its shear component. The delayed phase transformation can also be explained by noting that the Martensite phase minimizes the internal energy whereas austenite maximizes the entropy.\cite{Elliott2013} Indeed, introducing vacancies increases the internal energy and the configurational entropy\cite{Kastner2006} favoring austenite over Martensite.  As a result, a greater energetic driving force is required for the transformation to occur.

\begin{figure} 
\includegraphics[width=3in]{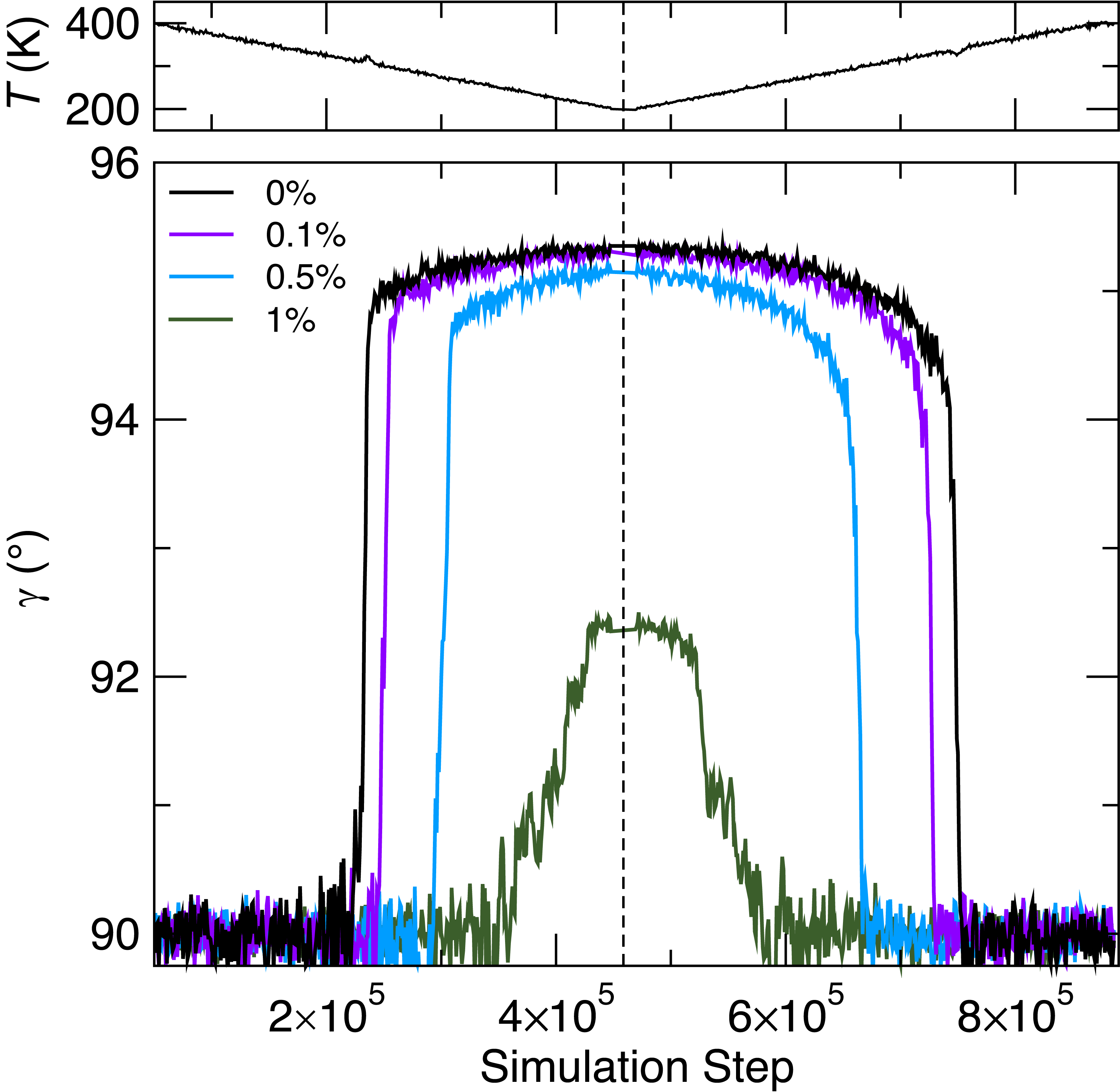} 
\caption{The change of unit cell angle ($\gamma$) with simulation step and temperature (shown above) with 0\%, 0.1\%, 0.5\%, and 1\% of anti-site defects indicates a large change in the structural transformation temperature. } 
\label{fig:NiTi-degree-a}
\end{figure}

NiTi with anti-site defect shows a similar change in $\gamma$ versus temperature (simulation step), Fig.\,\ref{fig:NiTi-degree-a}. In the temperature range of studied here, systems with 2\% and 3\% of anti-site defects did not show any phase transformation and likely occur at much lower temperature (\textless 200\,K) than calculated. Increasing the concentration of anti-site defects, the same modification of the Martensitic transformation occurs.   The higher concentration of defects forces the transformation to occur at lower temperature and result in a smaller unit cell angle.  These changes are even more pronounced than in the vacancy model with the transformation temperature with 1\% of anti-site defects resulting in similar properties to the 3\% of vacancy model.  Moreover, $\gamma$ in B19$^\prime$ is dramatically suppressed with an angle of $\approx$92.5$^\circ$ for the 1\% anti-site calculation compared to 95.2$^\circ$ for the 1\% vacancy.

Research has shown that a slight change in the composition of NiTi forming Ni-rich or Ti-rich phases also affects the phase transition drastically.\cite{Waitz2005, Frenzel2010, Tang1999} For example, the addition of $\approx$1\% excess Ni causes M$s$ to decreases by 100\,K.\cite{Tang1999} The average transformation temperature for the 1\% anti-site defect structure calculated here is 81\,K lower than the perfect structure.  This phenomenon can be rationalized by noting that increasing the concentration of Ni or Ti produces elemental islands (eith FCC or HCP) in the matrix. These islands energetically prefer to remain in the B2 (austenite) crystal structure rather than undergo the transformation.\cite{Saitoh2006}  Because the anti-site models used here mimic the formation of elemental Ni and Ti nano-island, the  suppression of transformation temperature is consistent.
 
\begin{figure} 
\includegraphics[width=3in]{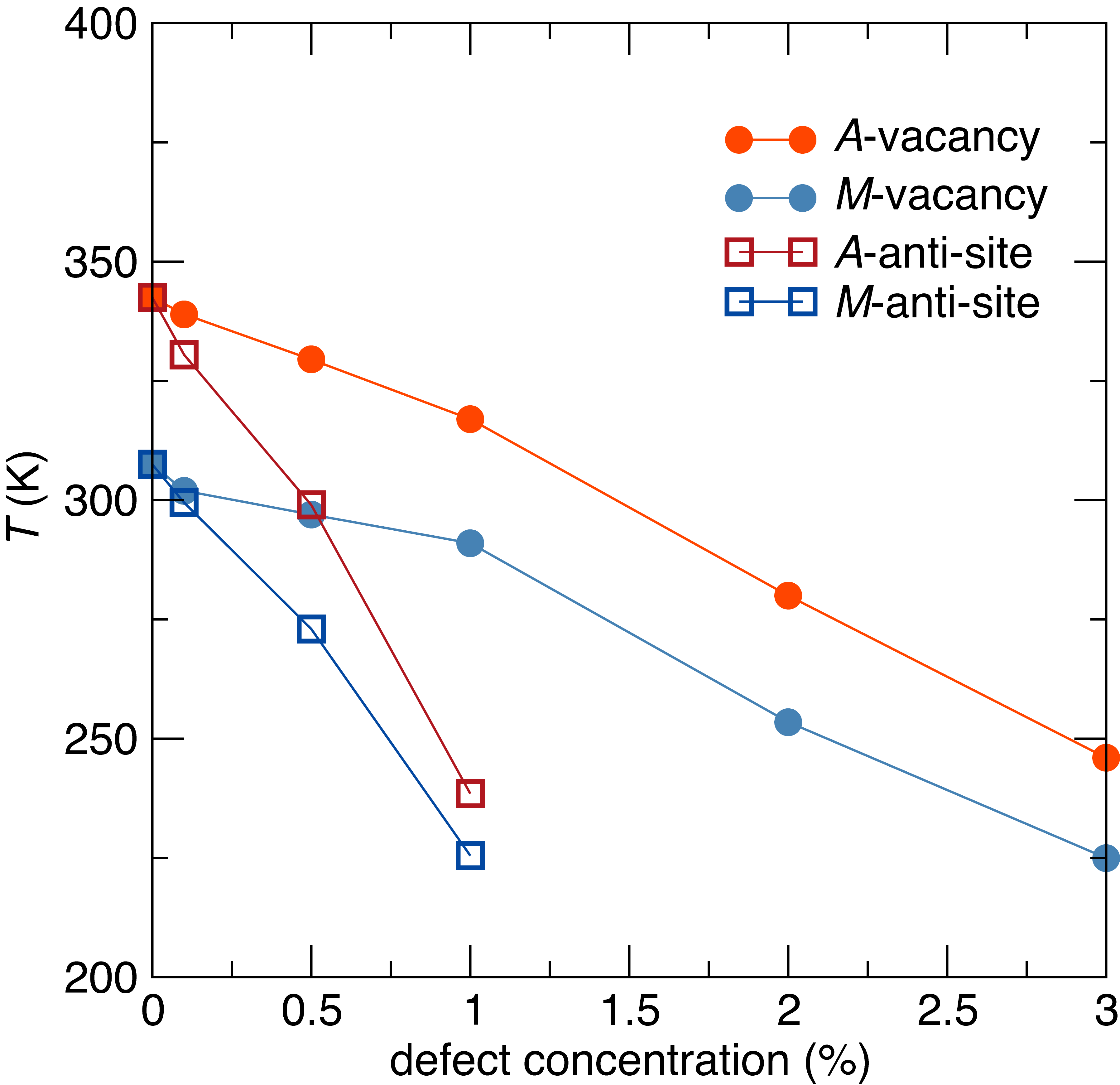} 
\caption{Changes of unit cell angle with simulation step and temperature with various percentages of vacancies (closed circles) and anti-site (open squares) defects.  The Martensitic and austenite transformations temperature are determined from the average of the start and finish Martensitic and austenite temperature respectively.} 
\label{fig:tva}
\end{figure}

The changes in the transformation temperatures with the incorporation of vacancy and anti-site defects are summarized in Fig.\,\ref{fig:tva}  The average of the start and finish temperatures are plotted against the point defect concentration.  The transformation occurs at much lower concentrations of anti-site defects compared to vacancies.  This analysis shows that at 1\% defect concentration, the anti-site defects reduce the transformation temperature by $\approx$26\% for the cooling cycle and $\approx$19.5\% for the heating cycle compared to the vacancy model.   Furthermore, the hysteresis of the transformation also decreases with an increasing concentration of point defects.

\section{Conclusion}

Molecular dynamic simulations are a valuable tool to study the effect of point defects on the phase transformation of NiTi during cooling and heating. Both vacancy and anti-site defects yield lower phase transformation temperatures and angles. The Martensitic transformation changes due to pinning effects from the point defects. In addition, systems containing anti-site defects further resist to the transformation due to the formation of nano-islands of the pure metals which favor the austenite phase.  
The study of the point defect within this phase transformation not only provides insight into the physics of this phase transformation but it also enables more fine-tune adjustment of the transformation temperatures through defect chemistry.

\section{Acknowledgments}

The authors thank the Department of Chemistry and the Division of Research at the University of Houston for providing generous start-up funds.  
This work was also supported by the R. A. Welch Foundation and the State of Texas through the Texas
Center for Superconductivity.
The research presented here used the Maxwell/Opuntia Cluster(s) operated by the University of Houston Center of Advanced Computing and Data Systems (CACDS).

\bibliography{NiTi}

\begin{thebibliography}{35}%
\makeatletter
\providecommand \@ifxundefined [1]{%
 \@ifx{#1\undefined}
}%
\providecommand \@ifnum [1]{%
 \ifnum #1\expandafter \@firstoftwo
 \else \expandafter \@secondoftwo
 \fi
}%
\providecommand \@ifx [1]{%
 \ifx #1\expandafter \@firstoftwo
 \else \expandafter \@secondoftwo
 \fi
}%
\providecommand \natexlab [1]{#1}%
\providecommand \enquote  [1]{``#1''}%
\providecommand \bibnamefont  [1]{#1}%
\providecommand \bibfnamefont [1]{#1}%
\providecommand \citenamefont [1]{#1}%
\providecommand \href@noop [0]{\@secondoftwo}%
\providecommand \href [0]{\begingroup \@sanitize@url \@href}%
\providecommand \@href[1]{\@@startlink{#1}\@@href}%
\providecommand \@@href[1]{\endgroup#1\@@endlink}%
\providecommand \@sanitize@url [0]{\catcode `\\12\catcode `\$12\catcode
  `\&12\catcode `\#12\catcode `\^12\catcode `\_12\catcode `\%12\relax}%
\providecommand \@@startlink[1]{}%
\providecommand \@@endlink[0]{}%
\providecommand \url  [0]{\begingroup\@sanitize@url \@url }%
\providecommand \@url [1]{\endgroup\@href {#1}{\urlprefix }}%
\providecommand \urlprefix  [0]{URL }%
\providecommand \Eprint [0]{\href }%
\providecommand \doibase [0]{http://dx.doi.org/}%
\providecommand \selectlanguage [0]{\@gobble}%
\providecommand \bibinfo  [0]{\@secondoftwo}%
\providecommand \bibfield  [0]{\@secondoftwo}%
\providecommand \translation [1]{[#1]}%
\providecommand \BibitemOpen [0]{}%
\providecommand \bibitemStop [0]{}%
\providecommand \bibitemNoStop [0]{.\EOS\space}%
\providecommand \EOS [0]{\spacefactor3000\relax}%
\providecommand \BibitemShut  [1]{\csname bibitem#1\endcsname}%
\let\auto@bib@innerbib\@empty
\bibitem [{\citenamefont {Otsuka}, \citenamefont {Sawamura},\ and\
  \citenamefont {Shimizu}(1971)}]{Otsuka1971}%
  \BibitemOpen
  \bibfield  {author} {\bibinfo {author} {\bibfnamefont {K.}~\bibnamefont
  {Otsuka}}, \bibinfo {author} {\bibfnamefont {T.}~\bibnamefont {Sawamura}}, \
  and\ \bibinfo {author} {\bibfnamefont {K.}~\bibnamefont {Shimizu}},\ }\href
  {\doibase 10.1002/pssa.2210050220} {\bibfield  {journal} {\bibinfo  {journal}
  {Phys.Status.Solidi.A}\ }\textbf {\bibinfo {volume} {5}},\ \bibinfo {pages}
  {457} (\bibinfo {year} {1971})}\BibitemShut {NoStop}%
\bibitem [{\citenamefont {Wagner}\ and\ \citenamefont
  {Windl}(2008)}]{Wagner2008}%
  \BibitemOpen
  \bibfield  {author} {\bibinfo {author} {\bibfnamefont {M.-X.}\ \bibnamefont
  {Wagner}}\ and\ \bibinfo {author} {\bibfnamefont {W.}~\bibnamefont {Windl}},\
  }\href {\doibase http://dx.doi.org/10.1016/j.actamat.2008.08.043} {\bibfield
  {journal} {\bibinfo  {journal} {Acta Mater.}\ }\textbf {\bibinfo {volume}
  {56}},\ \bibinfo {pages} {6232 } (\bibinfo {year} {2008})}\BibitemShut
  {NoStop}%
\bibitem [{\citenamefont {Hatcher}, \citenamefont {Kontsevoi},\ and\
  \citenamefont {Freeman}(2009)}]{Hatcher2009}%
  \BibitemOpen
  \bibfield  {author} {\bibinfo {author} {\bibfnamefont {N.}~\bibnamefont
  {Hatcher}}, \bibinfo {author} {\bibfnamefont {O.~Y.}\ \bibnamefont
  {Kontsevoi}}, \ and\ \bibinfo {author} {\bibfnamefont {A.~J.}\ \bibnamefont
  {Freeman}},\ }\href {\doibase 10.1103/PhysRevB.80.144203} {\bibfield
  {journal} {\bibinfo  {journal} {Phys. Rev. B}\ }\textbf {\bibinfo {volume}
  {80}},\ \bibinfo {pages} {144203} (\bibinfo {year} {2009})}\BibitemShut
  {NoStop}%
\bibitem [{\citenamefont {Holec}\ \emph {et~al.}(2011)\citenamefont {Holec},
  \citenamefont {Fri\'ak}, \citenamefont {Dlouh\'y},\ and\ \citenamefont
  {Neugebauer}}]{Holec2011}%
  \BibitemOpen
  \bibfield  {author} {\bibinfo {author} {\bibfnamefont {D.}~\bibnamefont
  {Holec}}, \bibinfo {author} {\bibfnamefont {M.}~\bibnamefont {Fri\'ak}},
  \bibinfo {author} {\bibfnamefont {A.}~\bibnamefont {Dlouh\'y}}, \ and\
  \bibinfo {author} {\bibfnamefont {J.}~\bibnamefont {Neugebauer}},\ }\href
  {\doibase 10.1103/PhysRevB.84.224119} {\bibfield  {journal} {\bibinfo
  {journal} {Phys. Rev. B}\ }\textbf {\bibinfo {volume} {84}},\ \bibinfo
  {pages} {224119} (\bibinfo {year} {2011})}\BibitemShut {NoStop}%
\bibitem [{\citenamefont {ichi Saitoh}, \citenamefont {Sato},\ and\
  \citenamefont {Shinke}(2006)}]{Saitoh2006}%
  \BibitemOpen
  \bibfield  {author} {\bibinfo {author} {\bibfnamefont {K.}~\bibnamefont {ichi
  Saitoh}}, \bibinfo {author} {\bibfnamefont {T.}~\bibnamefont {Sato}}, \ and\
  \bibinfo {author} {\bibfnamefont {N.}~\bibnamefont {Shinke}},\ }\href
  {\doibase 10.2320/matertrans.47.742} {\bibfield  {journal} {\bibinfo
  {journal} {Mater. Trans.}\ }\textbf {\bibinfo {volume} {47}},\ \bibinfo
  {pages} {742} (\bibinfo {year} {2006})}\BibitemShut {NoStop}%
\bibitem [{\citenamefont {Uehara}, \citenamefont {Asai},\ and\ \citenamefont
  {Ohno}(2009)}]{Uehara2009}%
  \BibitemOpen
  \bibfield  {author} {\bibinfo {author} {\bibfnamefont {T.}~\bibnamefont
  {Uehara}}, \bibinfo {author} {\bibfnamefont {C.}~\bibnamefont {Asai}}, \ and\
  \bibinfo {author} {\bibfnamefont {N.}~\bibnamefont {Ohno}},\ }\href@noop {}
  {\bibfield  {journal} {\bibinfo  {journal} {Modell. Simul. Mater. Sci. Eng.}\
  }\textbf {\bibinfo {volume} {17}},\ \bibinfo {pages} {035011} (\bibinfo
  {year} {2009})}\BibitemShut {NoStop}%
\bibitem [{\citenamefont {Mutter}\ and\ \citenamefont
  {Nielaba}(2010)}]{Mutter2010}%
  \BibitemOpen
  \bibfield  {author} {\bibinfo {author} {\bibfnamefont {D.}~\bibnamefont
  {Mutter}}\ and\ \bibinfo {author} {\bibfnamefont {P.}~\bibnamefont
  {Nielaba}},\ }\href@noop {} {\bibfield  {journal} {\bibinfo  {journal} {Phys.
  Rev. B}\ }\textbf {\bibinfo {volume} {82}},\ \bibinfo {pages} {224201}
  (\bibinfo {year} {2010})}\BibitemShut {NoStop}%
\bibitem [{\citenamefont {Shaw}\ and\ \citenamefont
  {Kyriakides}(1995)}]{Shaw1995}%
  \BibitemOpen
  \bibfield  {author} {\bibinfo {author} {\bibfnamefont {J.~A.}\ \bibnamefont
  {Shaw}}\ and\ \bibinfo {author} {\bibfnamefont {S.}~\bibnamefont
  {Kyriakides}},\ }\href {\doibase
  http://dx.doi.org/10.1016/0022-5096(95)00024-D} {\bibfield  {journal}
  {\bibinfo  {journal} {J. Mech. Phys. Solids}\ }\textbf {\bibinfo {volume}
  {43}},\ \bibinfo {pages} {1243 } (\bibinfo {year} {1995})}\BibitemShut
  {NoStop}%
\bibitem [{\citenamefont {Wang}, \citenamefont {Buehler},\ and\ \citenamefont
  {Pickart}(1965)}]{Wang1965}%
  \BibitemOpen
  \bibfield  {author} {\bibinfo {author} {\bibfnamefont {F.~E.}\ \bibnamefont
  {Wang}}, \bibinfo {author} {\bibfnamefont {W.~J.}\ \bibnamefont {Buehler}}, \
  and\ \bibinfo {author} {\bibfnamefont {S.~J.}\ \bibnamefont {Pickart}},\
  }\href {\doibase http://dx.doi.org/10.1063/1.1702955} {\bibfield  {journal}
  {\bibinfo  {journal} {J. Appl. Phys.}\ }\textbf {\bibinfo {volume} {36}},\
  \bibinfo {pages} {3232} (\bibinfo {year} {1965})}\BibitemShut {NoStop}%
\bibitem [{\citenamefont {Frenzel}\ \emph {et~al.}(2010)\citenamefont
  {Frenzel}, \citenamefont {George}, \citenamefont {Dlouhy}, \citenamefont
  {Somsen}, \citenamefont {Wagner},\ and\ \citenamefont
  {Eggeler}}]{Frenzel2010}%
  \BibitemOpen
  \bibfield  {author} {\bibinfo {author} {\bibfnamefont {J.}~\bibnamefont
  {Frenzel}}, \bibinfo {author} {\bibfnamefont {E.}~\bibnamefont {George}},
  \bibinfo {author} {\bibfnamefont {A.}~\bibnamefont {Dlouhy}}, \bibinfo
  {author} {\bibfnamefont {C.}~\bibnamefont {Somsen}}, \bibinfo {author}
  {\bibfnamefont {M.-X.}\ \bibnamefont {Wagner}}, \ and\ \bibinfo {author}
  {\bibfnamefont {G.}~\bibnamefont {Eggeler}},\ }\href {\doibase
  http://dx.doi.org/10.1016/j.actamat.2010.02.019} {\bibfield  {journal}
  {\bibinfo  {journal} {Acta Mater.}\ }\textbf {\bibinfo {volume} {58}},\
  \bibinfo {pages} {3444 } (\bibinfo {year} {2010})}\BibitemShut {NoStop}%
\bibitem [{\citenamefont {Khalil-Allafi}, \citenamefont {Dlouhy},\ and\
  \citenamefont {Eggeler}(2002)}]{Khalil-Allafi2002}%
  \BibitemOpen
  \bibfield  {author} {\bibinfo {author} {\bibfnamefont {J.}~\bibnamefont
  {Khalil-Allafi}}, \bibinfo {author} {\bibfnamefont {A.}~\bibnamefont
  {Dlouhy}}, \ and\ \bibinfo {author} {\bibfnamefont {G.}~\bibnamefont
  {Eggeler}},\ }\href {\doibase
  http://dx.doi.org/10.1016/S1359-6454(02)00257-4} {\bibfield  {journal}
  {\bibinfo  {journal} {Acta Mater.}\ }\textbf {\bibinfo {volume} {50}},\
  \bibinfo {pages} {4255 } (\bibinfo {year} {2002})}\BibitemShut {NoStop}%
\bibitem [{\citenamefont {Huang}, \citenamefont {Ackland},\ and\ \citenamefont
  {Rabe}(2003)}]{Huang2003}%
  \BibitemOpen
  \bibfield  {author} {\bibinfo {author} {\bibfnamefont {X.}~\bibnamefont
  {Huang}}, \bibinfo {author} {\bibfnamefont {G.~J.}\ \bibnamefont {Ackland}},
  \ and\ \bibinfo {author} {\bibfnamefont {K.~M.}\ \bibnamefont {Rabe}},\
  }\href@noop {} {\bibfield  {journal} {\bibinfo  {journal} {Nat. Mater.}\
  }\textbf {\bibinfo {volume} {2}},\ \bibinfo {pages} {307} (\bibinfo {year}
  {2003})}\BibitemShut {NoStop}%
\bibitem [{\citenamefont {Sanati}, \citenamefont {Albers},\ and\ \citenamefont
  {Pinski}(1998)}]{Sanati1998}%
  \BibitemOpen
  \bibfield  {author} {\bibinfo {author} {\bibfnamefont {M.}~\bibnamefont
  {Sanati}}, \bibinfo {author} {\bibfnamefont {R.~C.}\ \bibnamefont {Albers}},
  \ and\ \bibinfo {author} {\bibfnamefont {F.~J.}\ \bibnamefont {Pinski}},\
  }\href {\doibase 10.1103/PhysRevB.58.13590} {\bibfield  {journal} {\bibinfo
  {journal} {Phys. Rev. B}\ }\textbf {\bibinfo {volume} {58}},\ \bibinfo
  {pages} {13590} (\bibinfo {year} {1998})}\BibitemShut {NoStop}%
\bibitem [{\citenamefont {Lutton}, \citenamefont {Sabochick},\ and\
  \citenamefont {Lam}(1990)}]{Lutton1990}%
  \BibitemOpen
  \bibfield  {author} {\bibinfo {author} {\bibfnamefont {R.~T.}\ \bibnamefont
  {Lutton}}, \bibinfo {author} {\bibfnamefont {M.~J.}\ \bibnamefont
  {Sabochick}}, \ and\ \bibinfo {author} {\bibfnamefont {N.~Q.}\ \bibnamefont
  {Lam}},\ }in\ \href {\doibase 10.1557/PROC-209-207} {\emph {\bibinfo
  {booktitle} {Symposium K - Defects in Materials}}},\ \bibinfo {series} {MRS
  Online Proceedings Library}, Vol.\ \bibinfo {volume} {209}\ (\bibinfo {year}
  {1990})\BibitemShut {NoStop}%
\bibitem [{\citenamefont {Weber}, \citenamefont {Ablekim},\ and\ \citenamefont
  {Lynn}(2014)}]{Weber2014}%
  \BibitemOpen
  \bibfield  {author} {\bibinfo {author} {\bibfnamefont {M.~H.}\ \bibnamefont
  {Weber}}, \bibinfo {author} {\bibfnamefont {T.}~\bibnamefont {Ablekim}}, \
  and\ \bibinfo {author} {\bibfnamefont {K.~G.}\ \bibnamefont {Lynn}},\ }\href
  {http://stacks.iop.org/1742-6596/505/i=1/a=012006} {\bibfield  {journal}
  {\bibinfo  {journal} {J Phys Conf Ser.}\ }\textbf {\bibinfo {volume} {505}},\
  \bibinfo {pages} {012006} (\bibinfo {year} {2014})}\BibitemShut {NoStop}%
\bibitem [{\citenamefont {Lu}\ \emph {et~al.}(2007)\citenamefont {Lu},
  \citenamefont {Hu}, \citenamefont {Wang}, \citenamefont {Li}, \citenamefont
  {Xu},\ and\ \citenamefont {Yang}}]{Lu2007}%
  \BibitemOpen
  \bibfield  {author} {\bibinfo {author} {\bibfnamefont {J.~M.}\ \bibnamefont
  {Lu}}, \bibinfo {author} {\bibfnamefont {Q.~M.}\ \bibnamefont {Hu}}, \bibinfo
  {author} {\bibfnamefont {L.}~\bibnamefont {Wang}}, \bibinfo {author}
  {\bibfnamefont {Y.~J.}\ \bibnamefont {Li}}, \bibinfo {author} {\bibfnamefont
  {D.~S.}\ \bibnamefont {Xu}}, \ and\ \bibinfo {author} {\bibfnamefont
  {R.}~\bibnamefont {Yang}},\ }\href {\doibase 10.1103/PhysRevB.75.094108}
  {\bibfield  {journal} {\bibinfo  {journal} {Phys. Rev. B}\ }\textbf {\bibinfo
  {volume} {75}},\ \bibinfo {pages} {094108} (\bibinfo {year}
  {2007})}\BibitemShut {NoStop}%
\bibitem [{\citenamefont {Lai}\ and\ \citenamefont {Liu}(2000)}]{Lai2000}%
  \BibitemOpen
  \bibfield  {author} {\bibinfo {author} {\bibfnamefont {W.~S.}\ \bibnamefont
  {Lai}}\ and\ \bibinfo {author} {\bibfnamefont {B.~X.}\ \bibnamefont {Liu}},\
  }\href {http://stacks.iop.org/0953-8984/12/i=5/a=101} {\bibfield  {journal}
  {\bibinfo  {journal} {J. Phys.: Condens. Matter.}\ }\textbf {\bibinfo
  {volume} {12}},\ \bibinfo {pages} {L53} (\bibinfo {year} {2000})}\BibitemShut
  {NoStop}%
\bibitem [{\citenamefont {Kastner}\ \emph {et~al.}(2011)\citenamefont
  {Kastner}, \citenamefont {Eggeler}, \citenamefont {Weiss},\ and\
  \citenamefont {Ackland}}]{Kastner2011}%
  \BibitemOpen
  \bibfield  {author} {\bibinfo {author} {\bibfnamefont {O.}~\bibnamefont
  {Kastner}}, \bibinfo {author} {\bibfnamefont {G.}~\bibnamefont {Eggeler}},
  \bibinfo {author} {\bibfnamefont {W.}~\bibnamefont {Weiss}}, \ and\ \bibinfo
  {author} {\bibfnamefont {G.~J.}\ \bibnamefont {Ackland}},\ }\href {\doibase
  http://dx.doi.org/10.1016/j.jmps.2011.05.009} {\bibfield  {journal} {\bibinfo
   {journal} {J. Mech. Phys. Solids}\ }\textbf {\bibinfo {volume} {59}},\
  \bibinfo {pages} {1888 } (\bibinfo {year} {2011})}\BibitemShut {NoStop}%
\bibitem [{\citenamefont {Wu}, \citenamefont {Sung},\ and\ \citenamefont
  {Fang}(2013)}]{Wu2013}%
  \BibitemOpen
  \bibfield  {author} {\bibinfo {author} {\bibfnamefont {C.-D.}\ \bibnamefont
  {Wu}}, \bibinfo {author} {\bibfnamefont {P.-H.}\ \bibnamefont {Sung}}, \ and\
  \bibinfo {author} {\bibfnamefont {T.-H.}\ \bibnamefont {Fang}},\ }\href
  {\doibase 10.1007/s00894-013-1752-9} {\bibfield  {journal} {\bibinfo
  {journal} {J. Mol. Model.}\ }\textbf {\bibinfo {volume} {19}},\ \bibinfo
  {pages} {1883} (\bibinfo {year} {2013})}\BibitemShut {NoStop}%
\bibitem [{\citenamefont {Plimpton}(1995)}]{Plimpton1995}%
  \BibitemOpen
  \bibfield  {author} {\bibinfo {author} {\bibfnamefont {S.}~\bibnamefont
  {Plimpton}},\ }\href@noop {} {\bibfield  {journal} {\bibinfo  {journal} {J.
  Comput. Phys.}\ }\textbf {\bibinfo {volume} {117}},\ \bibinfo {pages} {1 }
  (\bibinfo {year} {1995})}\BibitemShut {NoStop}%
\bibitem [{\citenamefont {Finnis}\ and\ \citenamefont
  {Sinclair}(1984)}]{Finnis1984}%
  \BibitemOpen
  \bibfield  {author} {\bibinfo {author} {\bibfnamefont {M.~W.}\ \bibnamefont
  {Finnis}}\ and\ \bibinfo {author} {\bibfnamefont {J.~E.}\ \bibnamefont
  {Sinclair}},\ }\href@noop {} {\bibfield  {journal} {\bibinfo  {journal}
  {Philos. Mag.}\ }\textbf {\bibinfo {volume} {50}},\ \bibinfo {pages} {45}
  (\bibinfo {year} {1984})}\BibitemShut {NoStop}%
\bibitem [{\citenamefont {Kastner}(2003)}]{Kastner2003}%
  \BibitemOpen
  \bibfield  {author} {\bibinfo {author} {\bibfnamefont {O.}~\bibnamefont
  {Kastner}},\ }\href {\doibase 10.1007/s00161-003-0128-2} {\bibfield
  {journal} {\bibinfo  {journal} {Contin. Mech. Thermodyn}\ }\textbf {\bibinfo
  {volume} {15}},\ \bibinfo {pages} {487} (\bibinfo {year} {2003})}\BibitemShut
  {NoStop}%
\bibitem [{\citenamefont {Xie~*}\ and\ \citenamefont {Liu}(2004)}]{Xie2004}%
  \BibitemOpen
  \bibfield  {author} {\bibinfo {author} {\bibfnamefont {Z.~L.}\ \bibnamefont
  {Xie~*}}\ and\ \bibinfo {author} {\bibfnamefont {Y.}~\bibnamefont {Liu}},\
  }\href@noop {} {\bibfield  {journal} {\bibinfo  {journal} {Philos. Mag.}\
  }\textbf {\bibinfo {volume} {84}},\ \bibinfo {pages} {3497} (\bibinfo {year}
  {2004})}\BibitemShut {NoStop}%
\bibitem [{\citenamefont {Hoover}(1985)}]{Hoover1985}%
  \BibitemOpen
  \bibfield  {author} {\bibinfo {author} {\bibfnamefont {W.~G.}\ \bibnamefont
  {Hoover}},\ }\href {\doibase 10.1103/PhysRevA.31.1695} {\bibfield  {journal}
  {\bibinfo  {journal} {Phys. Rev. A}\ }\textbf {\bibinfo {volume} {31}},\
  \bibinfo {pages} {1695} (\bibinfo {year} {1985})}\BibitemShut {NoStop}%
\bibitem [{\citenamefont {Stukowski}(2010)}]{Stukowski2010}%
  \BibitemOpen
  \bibfield  {author} {\bibinfo {author} {\bibfnamefont {A.}~\bibnamefont
  {Stukowski}},\ }\href {http://stacks.iop.org/0965-0393/18/i=1/a=015012}
  {\bibfield  {journal} {\bibinfo  {journal} {Modell. Simul. Mater. Sci. Eng.}\
  }\textbf {\bibinfo {volume} {18}},\ \bibinfo {pages} {015012} (\bibinfo
  {year} {2010})}\BibitemShut {NoStop}%
\bibitem [{\citenamefont {{P. Sittner}}\ \emph {et~al.}(2003)\citenamefont {{P.
  Sittner}}, \citenamefont {{P. Lukás}}, \citenamefont {{D. Neov}},
  \citenamefont {{V. Novák}},\ and\ \citenamefont {{D.M.
  Toebbens}}}]{SLNNT2003}%
  \BibitemOpen
  \bibfield  {author} {\bibinfo {author} {\bibnamefont {{P. Sittner}}},
  \bibinfo {author} {\bibnamefont {{P. Lukás}}}, \bibinfo {author}
  {\bibnamefont {{D. Neov}}}, \bibinfo {author} {\bibnamefont {{V. Novák}}}, \
  and\ \bibinfo {author} {\bibnamefont {{D.M. Toebbens}}},\ }\href {\doibase
  10.1051/jp4:2003981} {\bibfield  {journal} {\bibinfo  {journal} {J. Phys. IV
  France}\ }\textbf {\bibinfo {volume} {112}},\ \bibinfo {pages} {709}
  (\bibinfo {year} {2003})}\BibitemShut {NoStop}%
\bibitem [{\citenamefont {Pelton}, \citenamefont {Dicello},\ and\ \citenamefont
  {Miyazaki}(2000)}]{Pelton2000}%
  \BibitemOpen
  \bibfield  {author} {\bibinfo {author} {\bibfnamefont {A.~R.}\ \bibnamefont
  {Pelton}}, \bibinfo {author} {\bibfnamefont {J.}~\bibnamefont {Dicello}}, \
  and\ \bibinfo {author} {\bibfnamefont {S.}~\bibnamefont {Miyazaki}},\
  }\href@noop {} {\bibfield  {journal} {\bibinfo  {journal} {Minim. Invasive.
  Ther. Allied. Technol.}\ }\textbf {\bibinfo {volume} {9}},\ \bibinfo {pages}
  {107} (\bibinfo {year} {2000})}\BibitemShut {NoStop}%
\bibitem [{\citenamefont {Zhong}, \citenamefont {Gall},\ and\ \citenamefont
  {Zhu}(2012)}]{Zhong2012}%
  \BibitemOpen
  \bibfield  {author} {\bibinfo {author} {\bibfnamefont {Y.}~\bibnamefont
  {Zhong}}, \bibinfo {author} {\bibfnamefont {K.}~\bibnamefont {Gall}}, \ and\
  \bibinfo {author} {\bibfnamefont {T.}~\bibnamefont {Zhu}},\ }\href {\doibase
  http://dx.doi.org/10.1016/j.actamat.2012.08.004} {\bibfield  {journal}
  {\bibinfo  {journal} {Acta Mater.}\ }\textbf {\bibinfo {volume} {60}},\
  \bibinfo {pages} {6301 } (\bibinfo {year} {2012})}\BibitemShut {NoStop}%
\bibitem [{\citenamefont {Ishida}\ and\ \citenamefont
  {Hiwatari}(2007)}]{Ishida2007}%
  \BibitemOpen
  \bibfield  {author} {\bibinfo {author} {\bibfnamefont {H.}~\bibnamefont
  {Ishida}}\ and\ \bibinfo {author} {\bibfnamefont {Y.}~\bibnamefont
  {Hiwatari}},\ }\href@noop {} {\bibfield  {journal} {\bibinfo  {journal} {Mol.
  Simul.}\ }\textbf {\bibinfo {volume} {33}},\ \bibinfo {pages} {459} (\bibinfo
  {year} {2007})}\BibitemShut {NoStop}%
\bibitem [{\citenamefont {Khalil-Allafi}\ and\ \citenamefont
  {Amin-Ahmadi}(2009)}]{Khalil-Allafi2009}%
  \BibitemOpen
  \bibfield  {author} {\bibinfo {author} {\bibfnamefont {J.}~\bibnamefont
  {Khalil-Allafi}}\ and\ \bibinfo {author} {\bibfnamefont {B.}~\bibnamefont
  {Amin-Ahmadi}},\ }\href {\doibase
  http://dx.doi.org/10.1016/j.jallcom.2009.07.135} {\bibfield  {journal}
  {\bibinfo  {journal} {J. Alloys Compd.}\ }\textbf {\bibinfo {volume} {487}},\
  \bibinfo {pages} {363 } (\bibinfo {year} {2009})}\BibitemShut {NoStop}%
\bibitem [{\citenamefont {Patel}\ and\ \citenamefont
  {Cohen}(1953)}]{Patel1953}%
  \BibitemOpen
  \bibfield  {author} {\bibinfo {author} {\bibfnamefont {J.}~\bibnamefont
  {Patel}}\ and\ \bibinfo {author} {\bibfnamefont {M.}~\bibnamefont {Cohen}},\
  }\href {\doibase http://dx.doi.org/10.1016/0001-6160(53)90083-2} {\bibfield
  {journal} {\bibinfo  {journal} {Acta. Metall.}\ }\textbf {\bibinfo {volume}
  {1}},\ \bibinfo {pages} {531 } (\bibinfo {year} {1953})}\BibitemShut
  {NoStop}%
\bibitem [{\citenamefont {Elliott}\ and\ \citenamefont
  {Karls}(2013)}]{Elliott2013}%
  \BibitemOpen
  \bibfield  {author} {\bibinfo {author} {\bibfnamefont {R.~S.}\ \bibnamefont
  {Elliott}}\ and\ \bibinfo {author} {\bibfnamefont {D.~S.}\ \bibnamefont
  {Karls}},\ }\href {\doibase http://dx.doi.org/10.1016/j.jmps.2013.07.013}
  {\bibfield  {journal} {\bibinfo  {journal} {J. Mech. Phys. Solids}\ }\textbf
  {\bibinfo {volume} {61}},\ \bibinfo {pages} {2522 } (\bibinfo {year}
  {2013})}\BibitemShut {NoStop}%
\bibitem [{\citenamefont {Kastner}(2006)}]{Kastner2006}%
  \BibitemOpen
  \bibfield  {author} {\bibinfo {author} {\bibfnamefont {O.}~\bibnamefont
  {Kastner}},\ }\href {\doibase 10.1007/s00161-006-0016-7} {\bibfield
  {journal} {\bibinfo  {journal} {Contin. Mech. Thermodyn}\ }\textbf {\bibinfo
  {volume} {18}},\ \bibinfo {pages} {63} (\bibinfo {year} {2006})}\BibitemShut
  {NoStop}%
\bibitem [{\citenamefont {Waitz}\ \emph {et~al.}(2005)\citenamefont {Waitz},
  \citenamefont {Spišák}, \citenamefont {Hafner},\ and\ \citenamefont
  {Karnthaler}}]{Waitz2005}%
  \BibitemOpen
  \bibfield  {author} {\bibinfo {author} {\bibfnamefont {T.}~\bibnamefont
  {Waitz}}, \bibinfo {author} {\bibfnamefont {D.}~\bibnamefont {Spišák}},
  \bibinfo {author} {\bibfnamefont {J.}~\bibnamefont {Hafner}}, \ and\ \bibinfo
  {author} {\bibfnamefont {H.~P.}\ \bibnamefont {Karnthaler}},\ }\href
  {http://stacks.iop.org/0295-5075/71/i=1/a=098} {\bibfield  {journal}
  {\bibinfo  {journal} {EPL-EUROPHYS LETT}\ }\textbf {\bibinfo {volume} {71}},\
  \bibinfo {pages} {98} (\bibinfo {year} {2005})}\BibitemShut {NoStop}%
\bibitem [{\citenamefont {Tang}\ \emph {et~al.}(1999)\citenamefont {Tang},
  \citenamefont {Sundman}, \citenamefont {Sandström},\ and\ \citenamefont
  {Qiu}}]{Tang1999}%
  \BibitemOpen
  \bibfield  {author} {\bibinfo {author} {\bibfnamefont {W.}~\bibnamefont
  {Tang}}, \bibinfo {author} {\bibfnamefont {B.}~\bibnamefont {Sundman}},
  \bibinfo {author} {\bibfnamefont {R.}~\bibnamefont {Sandström}}, \ and\
  \bibinfo {author} {\bibfnamefont {C.}~\bibnamefont {Qiu}},\ }\href {\doibase
  http://dx.doi.org/10.1016/S1359-6454(99)00193-7} {\bibfield  {journal}
  {\bibinfo  {journal} {Acta Mater.}\ }\textbf {\bibinfo {volume} {47}},\
  \bibinfo {pages} {3457 } (\bibinfo {year} {1999})}\BibitemShut {NoStop}%
\end{thebibliography}%

\end{document}